\documentstyle[12pt]{article}
\input{epsfig}
\begin{document}

%\large

\title{Statistics of Resonances in a Semi-infinite  Disordered Chain }

\author{Herv\'{e} Kunz$^1$ and Boris Shapiro$^2$\\
$^1$Ecole Polytechnique F\'{e}d\'{e}rale de Lausanne,\\ SB-ITP, CH-1015 Lausanne, Switzerland\\
$^2$Technion-Israel Institute of
Technology\\Department of Physics, Haifa 32000, Israel}
\maketitle

\begin{abstract}
We study the average density of resonances (DOR) for a
semi-infinite disordered chain, coupled to the outside world by a
(semi-infinite) perfect lead. A set of equations is derived, which
provides the general framework for calculating the average DOR,
for an arbitrary disorder and coupling strength.  These general
equations are applied to the case of weak coupling  and an
asymptotically exact expression for the averaged DOR is derived,
in the limit of small resonance width. This expression is
universal, in the sense that it holds for any degree of disorder
and everywhere in the (unperturbed) energy band.
\end{abstract}

\vskip 2cm
\section{Introduction }
Open quantum systems often exhibit the phenomenon of resonances.
Resonances correspond to quasi-stationary states which have a long
life-time but eventually decay into the continuum. (A particle,
initially within the system, escapes to infinity.) One approach to
the problem of resonances is based on the study of the analytic
properties of the scattering matrix, $S(E)$, in the complex energy
plane.  Resonances correspond to the poles, $\tilde E_\alpha =
E_\alpha -{i\over 2}\Gamma_\alpha$, of $S(E)$  on the unphysical
sheet \cite{LL,Zeld}. An alternative approach amounts to solving
the stationary Schr\"odinger equation,  with the boundary
condition of an  outgoing wave only. This condition, which makes
the problem non-Hermitian, describes a particle ejected from the
system. The Schr\"odinger equation with such boundary condition
admits complex eigenvalues $\tilde E_\alpha$, which correspond to
the resonances \cite{LL,Zeld}. This kind of approach leads in a
natural way to a non-Hermitian effective Hamiltonian, and it has
been used for a long time in scattering theory, including
scattering in chaotic and disordered systems (\cite{ Datta, Kot1},
and references therein). There are many examples of resonances in
atomic and nuclear physics. More recently, there has been much
interest in resonant phenomena in the field of chaotic and
disordered systems (for a recent review see \cite{Kot1}).

There is  considerable amount of work concerning the distribution
$P(\Gamma)$ of  resonance widths in one-dimensional disordered
chains \cite{TEG, Comt, Orl, Tit, Kot2}. Numerical studies
presented in that work demonstrate that, in a broad range of
$\Gamma$, $P(\Gamma)\propto \Gamma^{-\gamma}$, with the exponent
$\gamma$ being close to 1. (A similar behavior  pertains also to
two- and three-dimensional systems with  localized states
\cite{Orl, Kot2}). An analytical calculation  \cite{Tit} has been
performed for a one-dimensional continuous (white-noise)
potential, in the semiclassical limit,  when the localization
length is large.  It has been shown in that work that in a broad
range of $\Gamma$, $P(\Gamma)$ is well fitted by a function
$\Gamma^{-1.25}$. For sufficiently small $\Gamma$, however,
$(1/\Gamma)$-behavior was obtained, followed by a sharp cutoff at
still smaller $\Gamma$, due to the finite size of the sample. A
different analytical approach was recently developed in
\cite{Kunz}, for a discrete tight-binding random chain. The limit
of strong disorder (i.e., opposite to that of Ref. \cite{Tit}) was
considered and the  $(1/\Gamma)$-behavior (for a semi-infinite
chain) was derived.

In the present paper we develop a new approach to the problem of
resonances. The approach is based on counting the number of poles
of a resolvent of the corresponding non-Hermitian Hamiltonian. We
derive a set of equations for a semi-infinite disordered  chain,
coupled to a (semi-infinite) lead. This set contains, in
principal, the full solution of the problem, for an arbitrary
coupling strength and disorder. The equations simplify
considerably in the weak coupling regime. For this case we
rigorously derive the asymptotically exact $(1/\Gamma)$-rule and
present a simple scaling formula, which contains only the product
of the localization length and the density of states, at the
relevant energy.
\section{The model and its Effective Hamiltonian}
The system is depicted in Fig.1.
%%%%%%%%%%%%%%%%%%%%%%
\vspace{24pt}
\par
\hspace{0.5in} \epsfbox{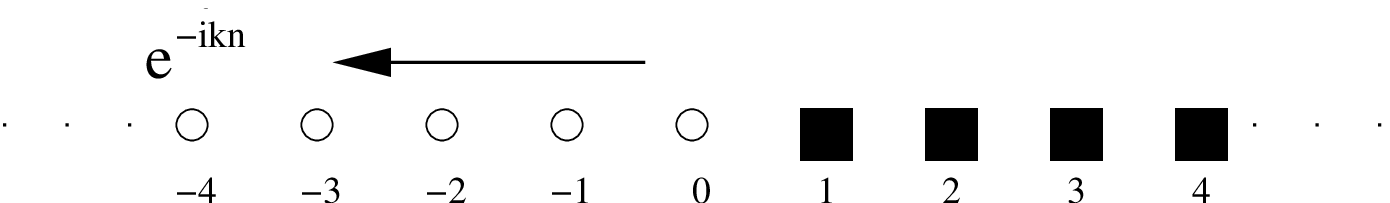}
\par
\baselineskip 12pt {\footnotesize {\bf Fig.~1:} A one-dimensional
chain coupled to a lead. Sites of the disordered chain are denoted
by black dots. Sites of the lead are denoted by open circles. The
arrow represents the out-going wave. \vspace{24pt}}
\par
\baselineskip 15pt
%%%%%%%%%%%%%%%%%%%%%
Black dots, labelled by $n=1,2,\ldots$, designate sites along the
semi-infinite disordered chain. Each site of the chain is assigned
a site energy, $\epsilon_n$. Different $\epsilon_n$'s
($n=1,2,\ldots$) are independent random variables chosen from some
distribution $q(\epsilon)$. Open circles, labeled by
$n=0,-1,-2,\ldots$ represent a perfect semi- infinite lead to which
the chain is coupled. All sites of the lead are assigned
$\epsilon_n = 0$. The lead simulates the free space outside the
chain. All nearest neighbor sites of the chain are coupled to each
other by a hopping amplitude $t$, and the same is true for all
nearest neighbor sites of the lead. The only exception to this
rule is the pair (0,1) which provides coupling between the chain
and the lead. The hopping amplitude for this pair is taken to be
equal $t'$. This allows us to tune the coupling from $t'=0$
(closed chain) to $t'=t$ (fully coupled chain).

%so that a particle, initially located somewhere within the chain,
%will eventually escape into the lead.

 As was mentioned in the Introduction, the most direct approach to
the problem of resonances is based on solving the stationary
Schr\"odinger equation  with the boundary condition of an {\em
outgoing wave} (see Fig.1). The Schr\"odinger equation for the
entire system (chain +lead) is a set of coupled equations:
\begin{eqnarray}\label{schrod1d}
-t\psi_{n+1} - t\psi_{n-1} &=& \tilde E \psi_n\quad\quad
\mbox{(for
$n<0$)}\\
-t\psi_{-1} - t'\psi_{1}  &=& \tilde E \psi_0\quad\quad(n=0)\\
-t\psi_{2} - t'\psi_{0} + \epsilon_1\psi_1 &=& \tilde E
\psi_1\quad\quad(n=1)\\
 -t\psi_{n+1} - t\psi_{n-1} + \epsilon_n\psi_n &=& \tilde
E \psi_n\quad\quad(n>1)
\end{eqnarray}
We recall that $\epsilon_n=0$ for $n<1$ (the lead) and it is
 random for $n\geq 1$ (the chain).
 Eqs.(1-4) are to be solved subjected to the boundary
condition of a outgoing wave in the lead, i.e. $\psi_n\propto \exp
(-i\tilde k n)$ , for $n\leq 0$, with $Im \tilde k <0$ . The complex wave vector $\tilde
k$ is related to $\tilde E$ by $\tilde E = -2t \cos \tilde k$.
Using the plane wave shape of the outgoing wave in the lead, it is
straightforward to eliminate from Eqs.(1-4) all $\psi_n$'s with
$n<1$ , thus reducing the problem to a system of equations for the
amplitudes $\psi_n$ on the sites of the disordered chain alone
($n=1,2,\ldots$):
\begin{equation}\label{nhschrod1d}
-t\psi_{n+1} - t\psi_{n-1} + \tilde\epsilon_n\psi_n = \tilde E
\psi_n\quad\quad (n = 1,2,\ldots  )\,
\end{equation}
with the condition $\psi_0=0$. Here $\tilde\epsilon_n =
\epsilon_n$ for $n=2,3, \ldots$, but not for $n=1$. This end site is
assigned  a complex energy
\begin{equation}\label{epsilon}
\tilde\epsilon_1 = \epsilon_1 -t\eta\exp i\tilde k  ,
\end{equation}
where the parameter $\eta =(t'/t)^2$ describes the coupling
strength to the outside world.
%(More generally,
%one could have here hopping amplitudes $t'$ different from $t$,
%which would allow for tuning of this coupling.)
 Thus, the
resonances are given by the complex eigenvalues of the
non-Hermitian effective Hamiltonian defined in (\ref{nhschrod1d}).
This non-Hermitian Hamiltonian, $\tilde H$, differs from the
Hermitian Hamiltonian, $H$, of the corresponding closed system
(i.e., with $\eta =0$) only by the complex correction to the
energy of the first site (Eq.(\ref{epsilon})):
\begin{equation}\label{H}
\tilde H = H -t\eta e^{i\tilde k}P  ,
\end{equation}
where $P$ is the projection on site 1. We set the energy scale by
taking $t=1$ and denote by $z$ the complex variable $\tilde E/t$.
Note that, since $\tilde k$ is related to $z$ via $z = -2 \cos
\tilde k$, the effective Hamiltonian $\tilde H$ depends on $z$.
Thus,  Eq. (\ref{nhschrod1d}) does not constitute a standard
eigen\-value problem and the eigenvalues of $\tilde H$ have to be
determined self-consistently. Defining formally the resolvent
$\tilde G(z)=(z-\tilde H)^{-1}$ we can write
\begin{equation}\label{resolvent} \tilde G = (z- H + \eta e^{
i\tilde k} P)^{-1}=[(z-H)(1+\eta e^ {i\tilde k} GP)]^{-1}=
 {1\over 1+\eta e^{ i\tilde k} GP}G   ,
\end{equation}
where $G=  (z-H)^{-1}$ is the "unperturbed" resolvent. Resonances
correspond to the singularities of the matrix $\tilde G_{nm}(z)$,
in the complex $z$-plane, or to the roots of the equation
\begin{equation}\label{roots}
1+\eta e^{ i\tilde k(z)} G_{11}(z)=0,
\end{equation}
where $G_{11}(z)$ is the $(1,1)$-element of $G$ in the site
representation.

\section{An expression for the average density of resonances}
Let us write  $G_{11}(z)$ as
\begin{equation}\label{selfenergy1}
{ G_{11}}(z) =  {1\over z- \epsilon_1 - S_1(z)},
\end{equation}
where $S_1(z)$ is the self-energy for site $1$. Eq.(\ref{roots})
can be written as
\begin{equation}\label{F}
F(z)\equiv z-\epsilon_1-S_1(z)+\eta e^{ i\tilde k(z)} =0.
\end{equation}
Denoting the zeroes of $F(z)$ by $z_\alpha$, we  write the density
of resonances (DOR), for a given realization of the disorder, as
$\rho(z)=\sum_{\alpha}\delta(z-z_\alpha)$. This expression refers
directly to the semi-infinite chain and it should be understood as
the  $ N\rightarrow \infty$ limit of the analogous expression for
a finite chain, of size $N$. The limit is well defined for any
$Imz\neq 0$ and no division of the sum by $N$ is necessary
\cite{Kunz}, in contrast to the usual case of the density of
states (on the real axis)  for a Hermitian problem.

 It is convenient to define a
$\delta$-function of a complex variable $F$ as $\delta(F)\equiv
\delta(ReF)\delta(ImF)$ and to use the relation
\begin{equation}\label{delta}
\sum_{\alpha}\delta(z-z_\alpha)=|F'(z)|^2\delta(F(z)),
\end{equation}
where the prime indicates a derivative with respect to $z$. This
relation is a generalization of the corresponding relation for a
real variable. It is valid if $F(z)$ is analytic in a domain containing the zeroes $z_\alpha$ and $F'(z_\alpha)\neq 0$. Indeed, under such conditions in the neighborhood of a zero, the equation $w=F(z)$ can be solved by an analytic function $z=z(w)$ and the jacobian of the change of variables from $(Re w,Im w)$ to $(Re z,Im z)$ is $|F'(z)|^2$ due to the Cauchy-Riemann equations. We will assume that $F'(z_\alpha)\neq0$ with probability one. Substituting  Eq.(\ref{F}) for $F(z)$, we have
\begin{equation}\label{rho}
\rho(z)=|1-S_1'(z)+\eta\lambda'(z)|^2\delta(z-\epsilon_1-S_1(z)+\eta\lambda(z)),
\end{equation}
where $\lambda(z)$ stands for the expression $e^{ i\tilde k(z)}$.

Next we average Eq.(\ref{rho})  over all realizations of the set
$\{\epsilon_n\}$ of the random site energies (we denote this
average by angular brackets).  Since the self-energy $S_1(z)$ does
not depend on the energy $\epsilon_1$ of the first site, we can
single out the variable  $\epsilon_1$ and average over it
explicitly. This leads to the following expression for the average
DOR:
\begin{equation}\label{rho1}
<\rho(z)>=\int\ d\epsilon q(\epsilon)<[|1+\eta\lambda'|^2
 -2Re((1+\eta\lambda')S'_1) +|S'_1|^2]\delta(z-\epsilon-S_1+\eta\lambda)>,
\end{equation}
The two correlated random variables, $S_1$ and $S'_1$ depend on
the set $\{\epsilon_2, \epsilon_3,...\}$ of the random site
energies. Note that we do not need to know the joint probability
distribution for $S_1$ and $S'_1$. Indeed, only three special
combinations of these two variables appear in Eq.(\ref{rho1}). It
is therefore useful to introduce three functions
\begin{eqnarray}\label{functions}
f(w)&=&<\delta(S_1-w)>\\
g(w)&=&<S_1'(z)\delta(S_1-w)>\\
h(w)&=&<|S_1'(z)|^2\delta(S_1-w)>
\end{eqnarray}
and write (\ref{rho1}) as
\begin{eqnarray}\label{rho2}
<\rho(z)>&=&\int\ d\epsilon q(\epsilon)[|1+\eta\lambda'(z)|^2
f(z-\epsilon +\eta\lambda(z))\nonumber\\
&&\hspace{10mm} -2Re(1+\eta\lambda'(z))g(z-\epsilon +\eta\lambda(z)) +h(z-\epsilon
 +\eta\lambda(z))].
\end{eqnarray}
We emphasize that $S_1$  refers   to a closed (semi-infinite)
chain, so that one can use the standard relations for various
quantities of the Hermitian problem \cite{LGP}. In particular,
we'll need the relation
\begin{equation}\label{recursion1}
S_1(z)= {1\over z-\epsilon_2-S_2(z)},
\end{equation}
where $S_2(z)$ is the self energy for site 2, with site 1
excluded. It immediately follows that
\begin{equation}\label{recursion2}
S'_1(z)= -S_1^2(z) (1-S_2'(z)).
\end{equation}

With the help of the definitions (15)-(17) and the recursion
relations (19), (20) it is straightforward to derive integral
relations between the functions $f$, $g$ and $h$. For instance,
using (20), $h(w)$ can be rewritten as
\begin{equation}\label{h1}
h(w)= <(1-2ReS_2'+|S_2'|^2)\delta(z-\epsilon_2 -S_2 -{1\over w})>.
\end{equation}
The variable $\epsilon_2$ is now "isolated" (i.e., all other
quantities in (\ref{h1}) do not depend on this variable) and can
be averaged over. This leads to
\begin{equation}\label{h2}
h(w)=\int\ d\epsilon q(\epsilon)[f(z-\epsilon-{1\over
w})-2Reg(z-\epsilon-{1\over w})+h(z-\epsilon-{1\over w})].
\end{equation}
The two other equations can be derived in a similar way, yielding:
\begin{equation}\label{g}
g(w)={w^2\over |w|^4}\int\ d\epsilon
q(\epsilon)[g(z-\epsilon-{1\over w})-f(z-\epsilon-{1\over w})]
\end{equation}
and
\begin{equation}\label{f}
f(w)={1\over |w|^4}\int\ d\epsilon q(\epsilon)f(z-\epsilon-{1\over
w}).
\end{equation}
Let us stress that the above defined functions of the complex
argument $w$ are not complex valued functions in the usual sense
but rather shorthand notations for a function of two real
variables, $w_1$ and $w_2$. For instance, Eq. (\ref{f}) for the
function $f(w)$, which is the probability distribution for the
real and imaginary part of the self-energy $S_1=w_1+iw_2$,  can be
written more explicitly as
\begin{equation}\label{f1}
f(w_1,w_2)={1\over(w_1^2+w_2^2)^2 }\int\ d\epsilon
q(\epsilon)f(x-\epsilon-{w_1\over w_1^2+w_2^2 },\quad y+{w_2\over
w_1^2+w_2^2 }).
\end{equation}

  The three equations, (22)-(24),
supplemented by the expression (18) provide the general framework
for computing the average DOR, $<\rho(z)>$, in the complex plane
$z=x+iy$, for any strength of disorder and for an arbitrary
coupling  $\eta$. We do not attempt to solve the problem in its
full generality but rather restrict ourselves to the small $\eta$,
i.e. weak coupling case.

\section{The weak coupling limit}
From now on we assume that the coupling constant $\eta$ is small
and develop  a "linear response theory" with respect to $\eta$. In
this limit the width of all resonances becomes proportional to
$\eta$, so that the "cloud" of resonances in the $(x,y)$-plane
gets squeezed towards the real axes. In order to define a
meaningful $ \eta\rightarrow 0$ limit for the DOR, one must
stretch the $y$-axis by a factor of $1/\eta$. It is natural to
switch from the original variable $z=x+iy$ to a new variable,
$Z=X+iY$, where $X=x$ and $Y=-y/\eta$, where the minus sign
accounts for the fact that the resonances are located in the lower
half plane of the complex variable $z$. Thus, in the $Z$-plane
resonances appear in the upper half-plane and $Y$ is the resonance
width, in units of $\eta$. The limiting (i.e., in the $
\eta\rightarrow 0$ limit) average DOR, $<\tilde\rho(X,Y)>$, in the
complex $Z$-plane is related to the original DOR, $<\rho(x,y)>$,
as:
\begin{equation}\label{DOR}
<\tilde\rho(X,Y)>=\lim_{\eta\rightarrow 0}\eta<\rho(X,-\eta Y)>.
\end{equation}

Next, we turn our attention to the functions $f, g$ and $h$ in the
$ \eta\rightarrow 0$ limit. The function $f(w_1,w_2)$ is the
probability distribution for the real and imaginary part of the
self-energy $S_1$. The shape of this function depends on the point
$z=x+iy$. In the $ \eta\rightarrow 0$ limit, as has been just
explained, one should set $y=-\eta Y$, where $Y$ does not depend
on $\eta$. Furthermore, since the imaginary part of $S_1$ is
proportional to $\eta$, we set $w_2=\eta W_2$. The distribution,
$F(W_1,W_2)$, for the new variables ($W_1=w_1, W_2=\eta^{-1}w_2)$
is related to the distribution $f(w_1,w_2)$ for the "old"
variables by $F(W_1,W_2)=\eta f(W_1, \eta W_2)$. The distribution
$F(W_1,W_2)$ has a well defined $ \eta\rightarrow 0$ limit and it
satisfies an integral equation, obtained from (\ref{f1}) by
transforming to the new variables and taking the $ \eta\rightarrow
0$ limit:
\begin{equation}\label{f2}
F(W_1,W_2)={1\over W_1^4}\int\ d\epsilon q(\epsilon)F(X-\epsilon
-{1\over W_1},\quad -Y + {W_2\over W_1^2}).
\end{equation}

The simplifying feature of the $ \eta\rightarrow 0$ limit is that
functions $g$ and $h$ are easily expressible in terms of $f$. More
precisely, we have to define functions $G$ and $H$, of the new
variables:
\begin{equation}\label{G}
G(W_1,W_2)=\eta g(W_1, \eta W_2),\quad H(W_1,W_2)=\eta h(W_1, \eta
W_2).
\end{equation}
These functions, in complete analogy with $F(W_1,W_2)$, have a
well defined $ \eta\rightarrow 0$ limit and satisfy integral
relations which are derived from (\ref{h2}), (\ref{g}) by writing
them in explicit form (compare to (\ref{f1})), transforming to the
new variables and  taking the $ \eta\rightarrow 0$ limit. This
yields:
\begin{eqnarray}\label{G1}
G(W_1,W_2)&=&{1\over W_1^2}\int\ d\epsilon q(\epsilon)[G(X-\epsilon
-{1\over W_1},\quad -Y + {W_2\over W_1^2})\quad\nonumber\\
&&\hspace{25mm}-F(X-\epsilon
-{1\over W_1},\quad -Y + {W_2\over W_1^2})],
\end{eqnarray}
and
\begin{eqnarray}\label{H1}
&&H(W_1,W_2)=\int\ d\epsilon q(\epsilon)[H(X-\epsilon -{1\over
W_1},\quad -Y + {W_2\over W_1^2})\quad \nonumber \\
&&\hspace{5mm}-2ReG(X-\epsilon -{1\over
W_1},\quad -Y + {W_2\over W_1^2})+
F(X-\epsilon -{1\over W_1},\quad -Y + {W_2\over W_1^2})].
\end{eqnarray}
%\begin{eqnarray}
%H(W_1,W_2)&=&\int\ d\epsilon q(\epsilon)\left[H(X-\epsilon -{1\over
%W_1},-Y + {W_2\over W_1^2})\right.\nonumber \\&&\left.-2ReG(X-\epsilon -{1\over
%W_1},-Y + {W_2\over W_1^2})+
%F(X-\epsilon -{1\over W_1},-Y + {W_2\over W_1^2})\right].
%\end{eqnarray}
One can easily  check that (29) and (30) are satisfied by
\begin{equation}\label{GH}
G(W_1,W_2)={W_2\over Y}F(W_1,W_2),\quad\quad H(W_1,W_2)=({W_2\over
Y})^2 F(W_1,W_2)
\end{equation}
Now we can derive an expression for $<\tilde\rho(X,Y)>$ $\langle\tilde\rho(X,y)\rangle$, in terms
of the function $F$, by taking the $ \eta\rightarrow 0$ limit in
Eq.(18) and using (\ref{DOR}) and (\ref{GH}):
\begin{equation}\label{rho3}
<\tilde\rho(X,Y)>= ({Im\lambda(X)\over Y})^2\int\ d\epsilon
q(\epsilon)F(X-\epsilon , -Y + Im\lambda(X)).
\end{equation}
Note that, since $z=X+i\eta Y$, in the $ \eta\rightarrow 0$ limit
$\tilde k(z)$ approaches $k(X)$ which is related to $X$ by
$X=-2\cos k$. Therefore $Im\lambda$ in (\ref{rho3}) is given by
\begin{equation}\label{lambda}
Im\lambda=\sqrt{1-{X^2\over 4}}.
\end{equation}
It is convenient to define a new variable, $J=W_2/Y$, so that $J$
is the imaginary part of self-energy (the resonance width) in
units of $Y\eta$. Finally, we denote $W_1$ by $R$ (the real part
of the self-energy) and define the probability distribution
$P(R,J)$, instead of $F(W_1,W_2)$. $P(R,J)$ satisfies the integral
equation
\begin{equation}\label{f2}
P(R,J)={1\over R^4}\int\ dR' q(x-{1\over R}-R')P(R', {J\over
R^2}-1),
\end{equation}
where the integration variable $\epsilon$ was replaced by
$R'=x-\epsilon -R^{-1}$.  The expression (\ref{rho3}) for the
average DOR now reads:
\begin{equation}\label{rho4}
<\tilde\rho(X,Y)>= {1\over Y^3}(1-{X^2\over 4})\int\ dR
q(X-R)P(R,\quad {1\over Y}\sqrt{1-{X^2\over 4}}-1).
\end{equation}
While performing integration in equations (\ref{f2}),
(\ref{rho4}), one should keep in mind that the function $P(R,J)$
is identically zero for $R>\sqrt J$. This property follows from
the basic recursion relation for the self energy $S$.

 The integral equation (\ref{f2}), supplemented by the
expression (\ref {rho4}), completely defines the problem of
resonances in the weak coupling limit. Note that the coupling
strength, $\eta$, does not appear explicitly in these equations.
It only determines the units in which the resonance width is
measured (the widths of all resonances is proportional to $\eta$).
Thus, $<\tilde\rho(X,Y)>$ is determined solely by the properties
of the closed system and, in this sense, Eqs.(\ref{f2}),
(\ref{rho4}) describe the regime of linear response with respect
to the coupling strength $\eta$. The function $P(R,J)$ describes
the joint probability distribution for the real and imaginary part
of the self-energy (of the closed system and for $z$ approaches
the real axis) and the integral equation (\ref{f2}) has appeared
previously in the study of Anderson localization \cite{Eco}.
Integration of (\ref{f2}) over $J$ yields the integral equation
for the probability distribution, $P(R)$, of the real part of the
self-energy:
\begin{equation}\label{f3}
P(R)={1\over R^2}\int\ dR' q(x-{1\over R}-R')P(R').
\end{equation}
This equation is very useful in the theory of one-dimensional
localization \cite{LGP,Kunz2}, because the knowledge of $P(R)$ allows
one to compute the localization length, $\xi$, according to:
\begin{equation}\label{f3}
{1\over \xi(X)}=\int\ dR'P(R')ln|R'|.
\end{equation}
We shall use this relation in the next Section.

\section{Large-$J$ asymptotics}

We were not able to obtain the complete analytic solution of the
integral equation (\ref{f2}). It is possible, however, to find the
large-$J$ asymptotics of the function $P(R,J)$. We state the
result and give the proof later: To leading order in the small
quantity, $1/J$,
\begin{equation}\label{large J}
P(R,J)={\nu(X)\xi(X)\over 2J^2},
\end{equation}
where $\nu(X)$ is the usual density of states, on the real energy
axis, for the closed chain. An important feature of this result is
that the large-$J$ asymptotics does not contain $R$. Therefore, by
plugging in (\ref{large J}) into (\ref{rho4}) we   immediately
obtain the simple  result for the small-$Y$ asymptotic of the
average DOR:
\begin{equation}\label{rho5}
<\tilde\rho(X,Y)>={\nu(X)\xi(X)\over 2Y}.
\end{equation}
Note that the factors containing $1-{X^2\over 4}\equiv D^2$ had
cancelled, so that  $X$ enters only via the density of states and
the localization length. This cancellation is the consequence of
the obvious scaling property of the general equation (\ref{rho4}),
namely: the dependence on $D$ drops out if one transforms to a new
variable, $Y\longrightarrow Y/D$, and to a new function,
$\rho\longrightarrow D\rho$. Thus, the asymptotic ($1/Y$)-
behavior is universal, in the sense that it holds for any degree
of disorder and for any $-2<X<2$.

It remains to prove the result stated in (\ref {large J}). It is
convenient to define the function
\begin{equation}\label{W}
W(R,J)=\int\limits_{J}^{\infty} dJ' P(R,J')
\end{equation}
and to study its large-$J$ asymptotics. An integral equation for
$W(R,J)$ is derived by integrating (\ref{f2}) over the second
argument of $P$. While performing this integration one should
remember that the function $P(R,J')$ is identically zero for
$J'<R^2$. It follows from this property that for $J<R^2$
\begin{equation}\label{W1}
W(R,J)=\int\limits_{0}^{\infty} dJ' P(R,J')=P(R).\quad\quad\quad
(J<R^2).
\end{equation}
Combining both cases, i.e., $J<R^2$ and $J>R^2$, one obtains the
following integral equation:
\begin{equation}\label{W2}
W(R,J)=\Theta(1-{J\over R^2})P(R)+ {1\over R^2}\Theta({J\over
R^2}-1)\int\ dR' q(x-{1\over R}-R')W(R', {J\over R^2}-1).
\end{equation}

We are interested in the large-$J$ behavior of $W(R,J)$. It is
useful to define the Laplace transform
\begin{equation}\label{W3}
\tilde W(R,s)=\int\limits_{0}^{\infty} dJe^{-sJ} W(R,J)
\end{equation}
and to study its small-$s$ behavior. The integral equation for the
transform is obtained directly from (\ref{W2}):

\begin{equation}\label{W4}
\tilde W(R,s)=\tilde W_0(R,s) + e^{-sR^2} \int\ dR' q(x-{1\over
R}-R')\tilde W(R', sR^2),
\end{equation}
where
\begin{equation}\label{W5}
\tilde W_0(R,s) ={1\over s}(1- e^{-sR^2})P(R).
\end{equation}
Anticipating that $W(R,J)$ is proportional to $1/J$, we look for
the solution (in the small $s$ limit) of (\ref{W4}) in the form
\begin{equation}\label{W6}
\tilde W(R,s)=A(R)\ln s+B(R)+\cdots
\end{equation}
This Ansatz satisfies Eq.(\ref {W4}) if
\begin{equation}\label{W7}
A(R)= \int\ dR' q(x-{1\over R}-R')A(R')
\end{equation}
and
\begin{equation}\label{W8}
B(R)= B_0(R) +\int\ dR' q(x-{1\over R}-R')B(R'),
\end{equation}
with
\begin{equation}\label{W9}
B_0(R)=R^2P(R)+A(R)\ln R^2.
\end{equation}
Since the site energy distribution, $q(\epsilon)$, is normalized
to 1, it immediately follows from (\ref{W7}) that $A(R)$ is a
constant. We denote this constant by $-a$ and determine it as
follows.

Define a function $\Phi(R')$ which is the non-trivial solution of
the integral equation
\begin{equation}\label{W10}
\Phi(R')= \int\ dR \Phi(R) q(x-{1\over R}-R').
\end{equation}
Note that the kernel $q(x-{1\over R}-R')$ is not symmetric with
respect to $R$ and $R'$ and that equation (\ref {W10}), unlike
(\ref {W7}), is not solved by a constant. The solution of (\ref
{W10}) is:
\begin{equation}\label{W11}
\Phi(R')={1\over R'^2}P({1\over R'}).
\end{equation}
We now go back to (\ref{W8}), multiply it by $\Phi(R)$, integrate
over $R$ and, with the help of (\ref{W10}), obtain:
\begin{equation}\label{W12}
\int\ dR \Phi(R)B_0(R)=0,
\end{equation}
which, using (\ref{W9}) and the aforementioned result
$A(R)=const=-a$, yields the value of $a$:
\begin{equation}\label{W13}
a={\int\ dR P(R)P({1\over R})\over \int\ dR {1\over R^2}P({1\over
R})\ln R^2},
\end{equation}
where the explicit form of $\Phi(R)$ (see (\ref{W11})) has been
used at the last step. The integral in the numerator is known in
localization theory \cite{Kunz2} and is equal to the
density of states $\nu(X)$. The integral in the denominator is
equal to $(2/\xi(X)$)(see (\ref{f3})). Thus, in the small-$s$
limit, $\tilde W(R,s)=-a\ln s$, with $a=\nu(X)\xi(X)/2$. As a
consequence, in the large-$J$ limit, $W(R,J)=a/J$ and
$P(R,J)=a/J^2$ which completes the proof of  (\ref{large J}). One can also show that (\ref{W8}) has a unique solution $B(R)$ and the behavior of $\tilde W(R,s)$ suggests that $W(R,J)=\alpha/J+R^2B(R)/J^2+\ldots$ and $p(R,J)=\alpha/J^2+2B(R)R^2/J^3+\ldots$

\section{Conclusions}
We studied the average density of resonances (DOR) for a
semi-infinite disordered chain, coupled to the outside world by a
(semi-infinite) perfect lead. The main result of this work is the
expression (18) for the average DOR, supplemented by the three
integral equations, (22-24), for the three functions, $h, g$ and
$f$. This set of equations provide the general framework for
calculating the average DOR, for an arbitrary disorder and
coupling strength. We applied these general equations to the case
of weak coupling and derived an asymptotically exact expression
for the average DOR, in the limit of small resonance width
(Eq.(39)). This expression is universal, in the sense that it
holds for any degree of disorder and everywhere in the
(unperturbed) energy band.

It is worthwhile to emphasize the essential difference between the
average DOR, as defined in this paper, and the probability
distribution of the resonance widths, often used in the literature
(e.g.\cite{TEG}). For a finite size chain, of $N$ sites, the two
quantities differ only by a factor $N$. However, when $N$
approaches $\infty$, the probability distribution shrinks towards
a delta-function, while the average DOR approaches a perfectly
well defined limit. Since we have taken a semi-infinite chain from
the start, it was essential for us to work with the average DOR,
rather than with the probability distribution of the resonant
widths. This enabled us to obtain the $1/Y$-behavior in the $
N\rightarrow \infty$ limit, in contrast to the statement of
\cite{TEG} that in that limit the $1/Y$-behavior has no region of
applicability.

The authors acknowledge useful conversations with E. Gurevich and
O. Kenneth.  One of the authors (B.S.) acknowledges the
hospitality of EPFL where most of this work was done. The work was
supported in part by the Israel Science Foundation (ISF).

H. Kunz thanks the Technion Institute for its generous hospitality.


\begin{thebibliography}{99}

\bibitem{LL}  L.D. Landau and E.M. Lifshitz, {\sl Quantum Mechanics \/}  (Pergamon, Oxford, 1977).

\bibitem{Zeld}  A.I. Baz, A. Perelomov and I.B. Zel'dovich,
{\sl Scattering, Reactions and Decay in Nonrelativistic Quantum
Mechanics\/} (Israel Program for Scientific Translations,
Jerusalem, 1969).

\bibitem{Datta} S. Datta, {\sl Electronic transport in mesoscopic systems\/}
(Cambridge University Press, Cambridge, 1995).

\bibitem{Kot1} T. Kottos, J. Phys. A {\bf 38}, 10761 (2005) (Special Issue on Trends in
Quantum Chaotic Scattering).

%\bibitem{Mir} A. D. Mirlin, Phys. Rep. {\bf 326}, 259 (2000).

\bibitem{TEG} M. Terraneo and I. Guarneri, Eur. Phys. J. {\bf B18}, 303 (2000).

\bibitem{Comt} C. Texier and A. Comtet, Phys. Rev. Lett. {\bf 82}, 4220,
(1999).

\bibitem{Orl} F.A. Pinheiro, M. Rusek, A. Orlowski and B.A. van Tiggelen,
 Phys. Rev. E {\bf 69}, 026605 (2004).

\bibitem{Tit} M. Titov and Y.V. Fyodorov,  Phys. Rev. B {\bf 61}, R2444 (2000).

\bibitem{Kot2} M. Weiss, J. A. Mendez-Bermudez and T. Kottos, Phys. Rev. B {\bf
73}, 045103 (2006).



\bibitem{Kunz} H. Kunz and B. Shapiro, J. Phys. A {\bf 39}, 10155
(2006).


%\bibitem{Eric} \'E. Akkermans and G. Montambaux, {\sl Physique M\'esoscopique
%des \'Electrons et des Photons\/} (EDP Sciences - CNRS \'Edition,
%2004).





%\bibitem{electrostatics}
%H.-J. Sommers, A. Crisanti, H. Sompolinski and Y. Stein, Phys.
%Rev. Lett. {\bf 60}, 1895 (1988).

%\bibitem{fz1} J. Feinberg and A. Zee,  Nucl. Phys. {\bf B504}, 579 (1997).

%\bibitem{Ginibre} J. Ginibre, {\sl Statistical Ensembles of Complex, Quaternion, and Real
%Matrices \/}, J. Math. Phys. {\bf 6}, 440 (1965).

\bibitem{LGP}  I.M. Lifshitz, S.A. Gredeskul and L.A. Pastur,
{\sl Introduction to the Theory of Disordered Systems\/} (Willey
and Sons, 1988).

%\bibitem{PW} P.W. Anderson, Phys. Rev. {\bf 109}, 1492 (1958).
\bibitem{Kunz2} H. Kunz and B. Souillard, Commun. Math. Phys. {\bf 79}, 201-249.
\bibitem{Eco}
C. Papatriantafillou, E. N. Economou and T. P. Eggarter, Phys.
Rev. B {\bf 13}, 910 (1976).



\end{thebibliography}
\end{document}